\documentclass{article}

\usepackage{arxiv}
\usepackage[version=3]{mhchem}
\usepackage[utf8]{inputenc} % allow utf-8 input
\usepackage[T1]{fontenc}    % use 8-bit T1 fonts
\usepackage{hyperref}       % hyperlinks
\usepackage{url}            % simple URL typesetting
\usepackage{booktabs}       % professional-quality tables
\usepackage{amsfonts}       % blackboard math symbols
\usepackage{nicefrac}       % compact symbols for 1/2, etc.
\usepackage{microtype}      % microtypography
\usepackage{lipsum}
\usepackage{graphicx}
\usepackage{epstopdf}
\usepackage{float}
\usepackage[caption = false]{subfig}
\usepackage{hyperref}
\newcommand{\DC}{$^\circ$C}

\title{Delayed nucleation in lipid particles}

\author{
  Guy Jacoby\\
  Department of Condensed Matter,\\
  The Raymond and Beverly Sackler\\ School of Physics and Astronomy,\\ 
  Tel-Aviv University,\\
  Ramat Aviv, Tel Aviv 6997801, Israel\\
  \texttt{guyjacob@tauex.tau.ac.il} \\
  \And
  Irina Portnaya \\
  Department of Chemical Engineering\\
  Technion-Israel Institute of Technology\\
  Haifa 3200003, Israel \\
  \And
  Dganit Danino \\
  Department of Chemical Engineering\\
  Technion-Israel Institute of Technology\\
  Haifa 3200003, Israel \\
  \And
  Haim Diamant \\
  The Raymond and Beverly School of Chemistry\\
  Tel-Aviv University,\\
  Ramat Aviv, Tel Aviv 6997801, Israel\\
  \And
  Roy Beck \\
  Department of Condensed Matter,\\
  The Raymond and Beverly Sackler\\ School of Physics and Astronomy,\\ 
  Tel-Aviv University,\\
  Ramat Aviv, Tel Aviv 6997801, Israel\\
  \texttt{roy@tauex.tau.ac.il} \\
}

\begin{document}
\maketitle

\begin{abstract}
Metastable states in first-order phase-transitions have been traditionally described by classical nucleation theory (CNT). However, recently an increasing number of systems displaying such a transition have not been successfully modelled by CNT. The delayed crystallization of phospholipids upon super-cooling is an interesting case, since the extended timescales allow access into the dynamics. Herein, we demonstrate the controllable behavior of the long-lived metastable liquid-crystalline phase of dilauroyl-phosphatidylethanolamine (DLPE), arranged in multi-lamellar vesicles, and the ensuing cooperative transition to the crystalline state. Experimentally, we find that the delay in crystallization is a bulk phenomenon, which is tunable and can be manipulated to span two orders of magnitude in time by changing the quenching temperature, solution salinity, or adding a secondary phospholipid. Our results reveal the robust persistence of the metastability, and showcase the apparent deviation from CNT. This distinctive suppression of the transition may be explained by the resistance of the multi-lamellar vesicle to deformations caused by nucleated crystalline domains. Since phospholipids are used as a platform for drug-delivery, a programmable design of cargo hold and release can be of great benefit.
\end{abstract}

\section{Introduction}
Classical nucleation theory (CNT), although decades old, is still the prevalent theory for many systems undergoing first-order phase-transitions. Only a few key assumptions are needed for CNT: dominant short range interactions, smooth interfaces (capillarity approximation), and that the nucleus has similar properties to the final bulk phase \cite{DeYoreo2003}. The dynamics are then described as a single stochastic excitation process that governs the transition. Mineral crystal formation \cite{Baumgartner2013a, Giuffre2013, Hamm2014}, virus capsid assembly \cite{Zandi2006}, and protein nucleation \cite{Akella2014, Sleutel2015} are examples of phase transition dynamics that can be successfully modeled by CNT. Despite its simplifying assumptions and basic description of the interactions, when applicable, CNT can properly capture the quantitative features of the nucleation process of many researched systems.

However, there is an increasing number of systems displaying nucleation processes that do not conform to this classical picture. Complex dynamics can arise due to intermediate states leading to multi-step nucleation \cite{Sleutel2014b, DeYoreo2015, Loh2017}, or long-range interactions that can result in macroscopic nucleation \cite{Nishino2011} or a cooperative delayed transition \cite{Neumann1985}. Such dynamics may require an extension of the classical theory or in some cases a comprehensive revision \cite{Chandra1989}. Examples of such complex dynamics can be found in self-assembled amphiphilic systems, which display long-lived metastable phases upon temperature change.

Amphiphiles are molecules that contain hydrophilic and hydrophobic chemical groups. The key characteristic associated with amphiphilic molecules is their ability to spontaneously self-assemble into macro-molecular structures \cite{Israelachvili2011}. Biological amphiphiles, such as lipids, self-assemble into a wide variety of mesophases. Most notably, lipids constitute the membranes of cells and organelles, and are involved in many important biological functions. Alongside basic research into their physical and biochemical properties, self-assembly is also utilized for designing modern biomedical applications such as drug delivery \cite{Pattni2015}.    

In particular, the phospholipid amphiphiles have several predominant lamellar phases, such as the disordered liquid-crystalline phase ($L_\alpha$), gel phase ($L_\beta$) and ordered crystalline phase ($L_c$), which differ by their degree of spatial symmetry. Transitions between these phases can be induced by changing the temperature, but the pathways depend on the physicochemical properties of the molecules and their thermal history. Previously, the phospholipid DLPE was shown to have long-lived metastable phases with lifetimes on the order of hours or even days \cite{Seddon1983, Chang1983, Jacoby2015}. These time-scales are orders of magnitude longer than the rapid transitions as in the case of melting. The reports were based mostly on X-ray scattering and differential scanning calorimetry (DSC), both very useful techniques for phase detection and characterization. However, they were performed as static measurements at different points in time, separated by long periods of unrecorded incubation. These limited observations only allowed for qualitative descriptions of the dynamics.

Herein, we present our experimental investigation of the metastable $L_\alpha$ to $L_{c}$ phase-transition using time-resolved solution X-ray scattering (SXS) and DSC measurements. We demonstrate the cooperative and controllable behavior of the transition dynamics. We highlight and discuss the deviations from CNT, and rationalize them based on the free-energy cost of deforming the $L_\alpha$ vesicles by crystal nucleation.

\section{Materials and methods}
\subsection{Lipid dispersion preparation}
1,2-dilauroyl-sn-glycero-3-phosphoethanolamine (DLPE), 1,2-dilauroyl-sn-glycero-3-phosphoglycerol (DLPG), 1,2-dimyristoyl-sn-glycero-3-phosphoglycerol (DMPG), 1,2-dipalmitoyl-sn-glycero-3-phosphophosphoglycerol (DPPG), 1,2-distearoyl-sn-glycero-3-phosphophosphoglycerol (DSPG) and 1,2-dilauroyl-sn-glycero-3-phosphocholine (DLPC) were purchased from Avanti Polar Lipids Inc. The lipids were dissolved in chloroform (DLPE) and chloroform:methanol 5:1 (other lipids) separately, then mixed together to achieve desired stoichiometry. Total lipid concentration was 30 mg/ml per sample. The solution was evaporated overnight in a fume hood, and re-fluidized using a buffer at 6.7 pH containing 20 mM 2-(N-Morpholino)ethanesulfonic acid (MES), 1 mM MgCl2 and 13 mM NaOH. NaCl was added to retain desired monovalent salt concentration. Samples were then placed in an incubator at 37 \DC{} for 3 hours, and homogenized using a vortexer every 25 minutes. Samples were then placed in quartz capillaries, containing about 100 $\mu l$, and centrifuged for 5 minutes at 3000 rpm, to create a pellet of lipids.

\subsection{Solution X-ray scattering}
Samples at 30 mg/ml lipid concentration were measured in 1.5 mm diameter sealed quartz capillaries. Measurements were performed using an in-house solution X-Ray scattering system, with a GeniX (Xenocs) low divergence $Cu~K_\alpha$ radiation source (wave length of 1.54 \AA) and a scatter-less slits setup \cite{Li2008}. Two-dimensional scattering data with a \textit{q} range of $0.06-2$ \AA{}$^{-1}$ at a sample-to-detector distance of about 230 mm were collected on a Pilatus 300K detector (Dectris), and radially integrated using MATLAB (MathWorks) based procedures (SAXSi). Background scattering data was collected from buffer solution alone. The background-subtracted scattering correlation peaks were fitted using a Gaussian with a linearly sloped baseline. For each sample, time-resolved correlation peaks position, intensity and width were extracted.

\subsection{Differential scanning calorimetry}
DSC experiments were performed using a VP-DSC micro calorimeter (MicroCal Inc., Northampton, MA). Calorimetric data analysis was done with the Origin 7.0 software. Degassed systems of pure DLPE and 90:10 DLPE:DLPG (mole \%) at a total concentration of 30 mg/ml were placed in the sample cell (0.5 ml), and MES buffer (20 mM MES + 130 mM NaCl at pH 6.7) in the reference cell. DSC thermograms were recorded during multiple heating-cooling cycles at various scan rates and different pre- and post-scan periods. First, each sample was heated from 25 \DC{} to 60 \DC{} with at a rate of 90 \DC{}/hour and 15 minute pre- and 3 hour post-scan periods. Then the sample was cooled from 60 \DC{} to 37 \DC{} with a scan rate of 90 \DC{}/hour, and identical 15 minute pre- and post-scan periods. The additional scans (presented in the text) were carried out at a very slow rate of 0.1 \DC{}/hour for heating and 0.43 \DC{}/hour for cooling, with 1 minute pre- and post-scan periods. In this manner, we were using the DSC in a quasi-isothermal mode, where we just wait for the metastability to end in an exothermic transition to the crystalline state. Additionally, on the same instrument DSC measurements of 10-fold diluted samples (3 mg/ml) of pure DLPE, and 97:3; 95:5; 92:8; 90:10, 85:15 DLPE:DLPG (mole \%) were performed, to measure the transition temperature and transition enthalpy. The buffer composition was the same. DSC thermograms were recorded during double heating-cooling cycles (between 25 \DC{}  and  60 \DC{}) with a scan rate of 90 \DC{}/hour, and identical 15 min pre- and post-scan periods. Calorimetric data analysis was done with the Origin 7.0.

\section{Results}
A fresh sample of DLPE in solution is found in the low energy $L_c$ phase at temperatures below 43 \DC{}. When heated above 43 \DC{}, the molecules undergo a melting transition $T_{c\rightarrow\alpha}$ to the $L_\alpha$ phase. \textit{Seddon et al.} showed that if the same sample is then cooled to a temperature $T_{Q}$ below $T_{c\rightarrow\alpha}$ there are two possible pathways back to the equilibrium $L_{c}$ phase \cite{Seddon1983}. If $T_{Q}<T_{\beta\rightarrow\alpha}$ (30 \DC{}), the gel-to-liquid crystalline transition, the system will rapidly transition to the $L_{\beta}$ phase, which will subsequently become metastable until returning to equilibrium. However, if $T_{\beta\rightarrow\alpha}<T_{Q}<T_{c\rightarrow\alpha}$, the $L_{\alpha}$ phase will become metastable and directly transition to the $L_{c}$ phase.

\begin{figure}[!htbp]
\centering
\subfloat[]{\includegraphics[width=0.5\columnwidth]{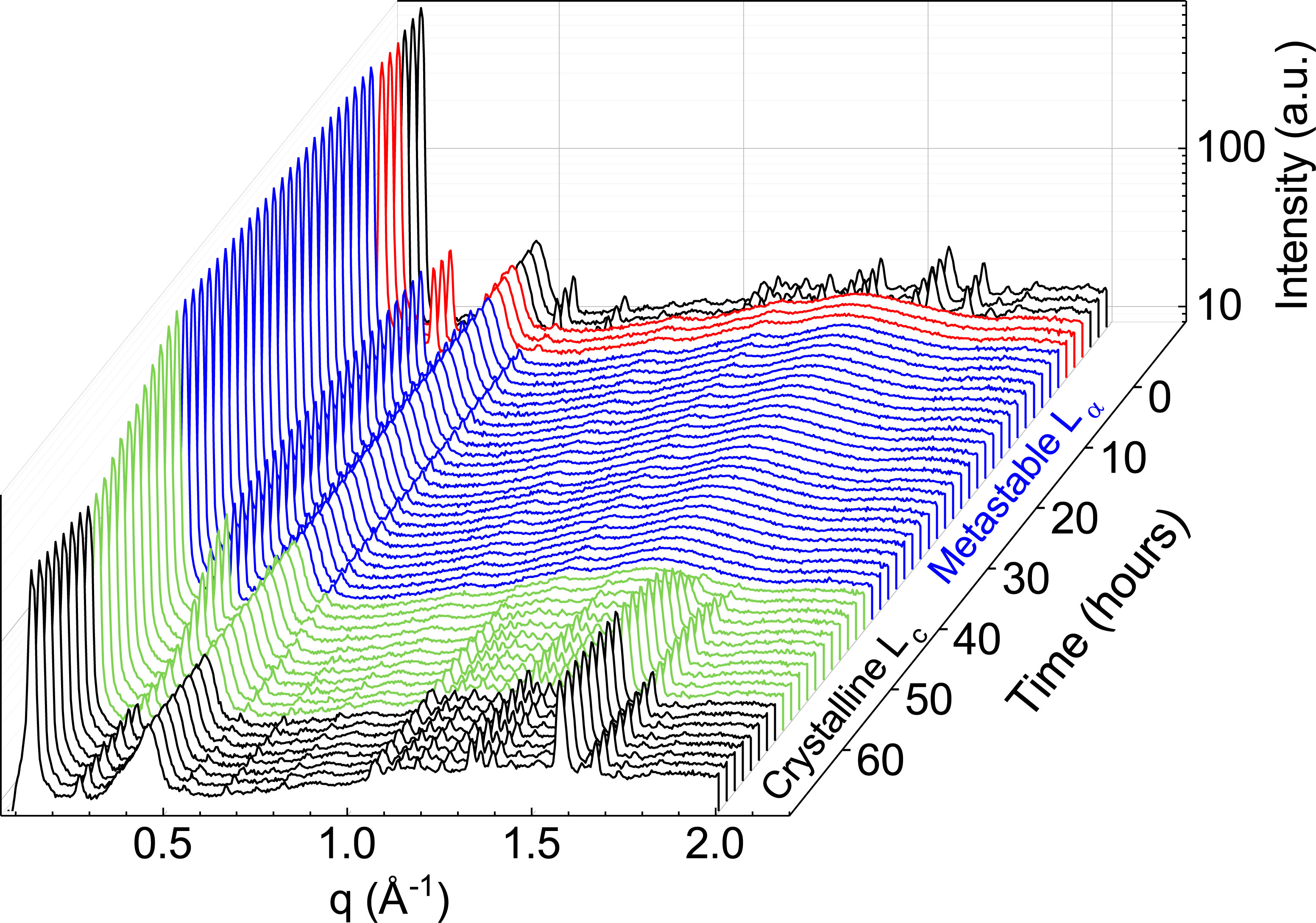}}
\subfloat[]{\includegraphics[width=0.5\columnwidth]{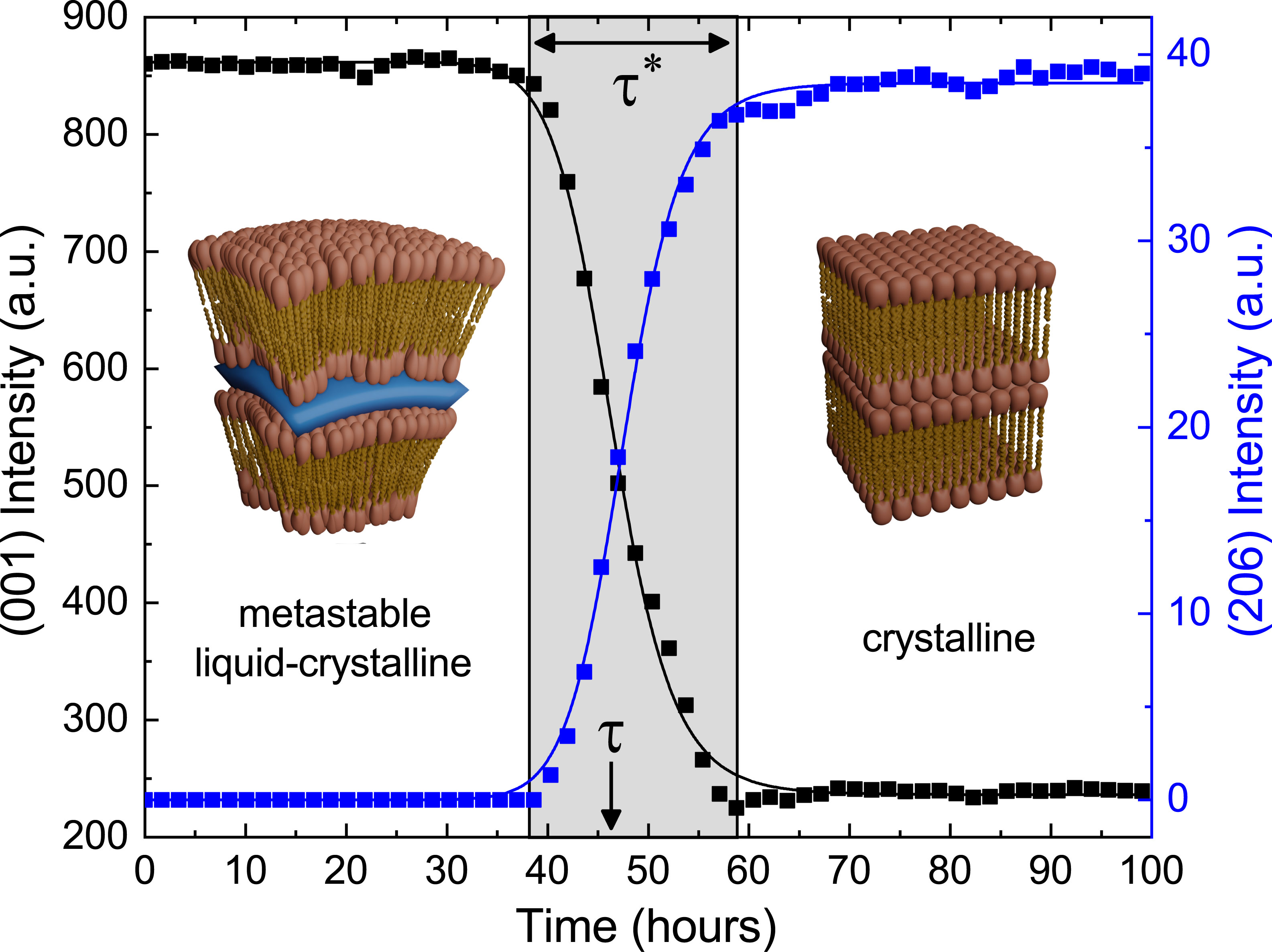}} 
\caption{(a) Time-resolved scattering spectra of lipid system containing 93:7 DLPE:DSPG (mole \%). Initially, at $T_{Q}=37$ \DC{} (black spectrum), the crystalline phase is indicated by the presence of wide-angle correlation peaks. When heated to 60 \DC{} (red spectra) these peaks disappear with the loss of in-plane order. After cooling back to $T_{Q}$ (blue spectra), the $L_{\alpha}$ phase remains metastable for $\tau=47$ hours, and then phase-transitions (green spectra) back into the crystalline phase (black spectra). (b) Scattering intensity of lamellar correlation peak (001) and mixed wide-angle peak (206) used to measure the order parameters and extract the temporal features of the dynamics. Inset: schematic illustration of the lipid conformation in both the liquid-crystalline ($L_{\alpha}$) and the crystalline ($L_{c}$) phases.}
\label{fgr:waterfall}
\end{figure}

Here, the metastable $L_{\alpha}$ to $L_{c}$ phase-transition was recorded by time-resolved SXS. In the experiments, the X-ray beam illuminating the sample has a cross sectional area of approximately 0.64 $mm^{2}$, which produces a bulk-averaged scattering signal. The scattering from a sample that has not been pre-heated is first recorded at $T_{Q}$ ($T_{\beta\rightarrow\alpha}<T_{Q}<T_{c\rightarrow\alpha}$) as a reference point to the initial low-energy state of the system. The scattering pattern of this initial state pertains to a 3D crystal with an orthorhombic unit cell, with the largest dimension corresponding to the lamellar repeating distance, and the shorter ones to the in-plane ordering of the lipids. The scattering from the sample is then recorded after heating to 60 \DC{} (above $T_{c\rightarrow\alpha}$), where the lipid membranes have transitioned to the smectic $L_{\alpha}$ phase, \textit{i.e.} a lamellar structure with liquid-like disorder in-plane. The phase change is accompanied by a change in morphology of the lipid particles, from faceted crystals to curved multi-lamellar vesicles (MLVs) \cite{Jacoby2015}. The in-plane order-to-disorder transition is marked by the disappearance of the correlation peaks at wide scattering angles (Fig. \ref{fgr:waterfall}).

After three hours at 60 \DC{}, the sample is quenched (at a rate of 2.5 \DC{}/min) back to $T_{Q}$ and measured every hour until it has finished transitioning back to $L_{c}$. The time-resolved scattering spectra show the evolution of the correlation peaks, which reports on structural changes and the phase-transition (Fig. \ref{fgr:waterfall}a). This heating-cooling procedure is the typical experiment conducted to record the metastability. Surprisingly, regardless of the parameters changed between samples, such as buffer salinity or the inclusion of a secondary lipid into the system, there are several prominent features in the evolution of the scattering spectrum after cooling back to $T_{Q}$. The first is the extended period of time in which the $L_{\alpha}$ phase remains metastable, which we denote as the delay time $\tau$ (Fig. \ref{fgr:waterfall}b). The second is the time the system spends transitioning between phases, denoted as the transition period $\tau^*$, which is by and large an order of magnitude shorter than $\tau$. The third notable feature occurring in most of the experiments accounted for in this work is a substantial change in the lamellar scattering intensity during the transition period. The intensity remains constant during the delay time, however it decreases, up to an order of magnitude, during the transition period. This indicates a decrease in the average number of lamellae per particle during the transition. This implies that the phase-transition is accompanied by a macro-scale structural change of the particles.

The delay time until crystallization ($\tau$) and the duration of the ensuing transition ($\tau^*$) are extracted from the time-dependent correlation peak intensities, which are a direct measurement of the order parameters. Specifically, we fit a sigmoid function to the time-dependent intensity of the lamellar correlation peak (001) and to a mixed correlation peak in the wide angles (206) (Fig. \ref{fgr:waterfall}b), to measure the out- and in-plane order parameters respectively.

In the classical nucleation theory, a single timescale should be observed, namely, that associated with the rate of nucleation. However, given that SXS produces bulk-averaged signals, one can immediately notice that the dynamics presented in the X-ray spectra do not readily conform to the classical picture of a single stochastic process culminating in a phase-transition. Instead of a gradual increase in crystalline scattering we observe two distinct timescales, $\tau$ and $\tau^*$, which are orders of magnitude larger than the melting transition times. Moreover, in all cases there is no detectable scattering at wide-angle prior to the transition, which implies a collective bulk transition rather than stochastic events of crystallization within the macroscopic illuminated area.

We set out to explore the different system parameters that can affect the dynamics of the transition. By changing system parameters such as the lipid stoichiometry and chemical structure, salinity and the quenching temperature ($T_{Q}$) we found that we were able to manipulate the metastability in a pre-determined and controllable fashion.

\begin{figure}[!htbp]
\centering
\includegraphics[width=0.5\columnwidth]{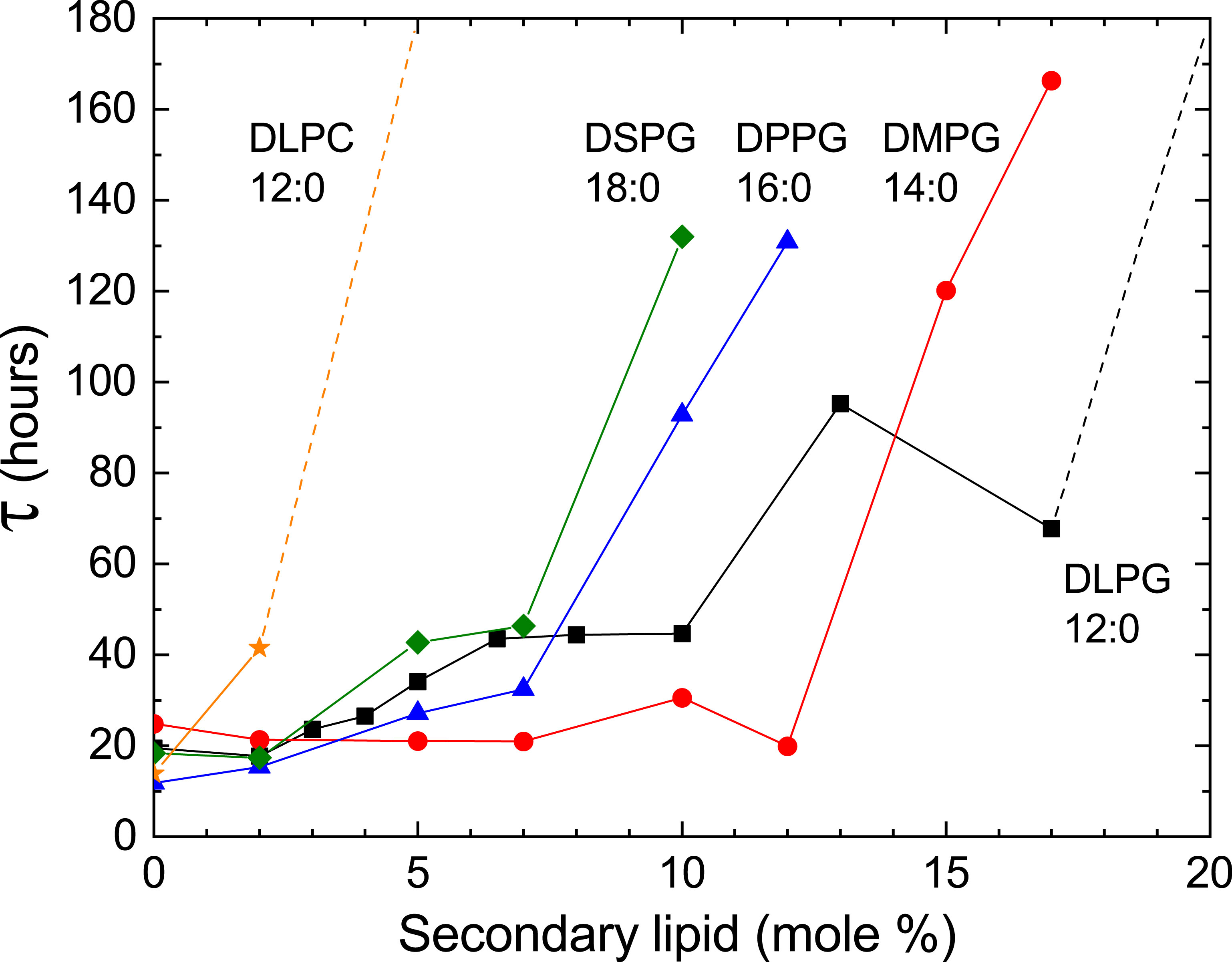}
\caption{Delay time ($\tau$) as a function of the molar fraction (mole \%) of the secondary lipid, labelled in the figure, \textit{i.e.} (100 - DLPE) (mole \%) . An increase in chain length results in an increase in delay time at a lower fraction. Dashed lines represent the minimal delay time for samples that did not transition within 180 hours. Samples with higher concentrations of DPPG and DSPG were measured but omitted from the results due to an alteration of the final crystalline form.}
\label{fgr:PGPC}
\end{figure}

In previous studies, the metastability was examined in samples of pure DLPE. However, we found that the addition of a secondary phospholipid not only preserves the metastable phase, but also extends its lifetime (Fig. \ref{fgr:PGPC}). Moreover, the delay time is sensitive to changes in the hydrocarbon chain length and headgroup. Phosphatidylglycerols (PGs) were chosen as a secondary charged lipid due to the stabilizing effect they have on PE bilayers \cite{Tari1989}, and specifically, DLPG was previously used along with DLPE as the building blocks for a drug delivery system \cite{Rivkin2010, Bachar2011, Cohen2014}. PGs with 12 (DLPG), 14 (DMPG), 16 (DPPG) or 18 (DSPG) carbons in their saturated hydrocarbon chains were chosen as chain length variants. In addition, the zwitterionic dilauroyl-phosphatidylcholine (DLPC) was chosen as a headgroup variant. The delay time seems to increase as a function of chain length for the PGs, and it is greatly increased when the PG headgroup is swapped with a PC.

\begin{figure}[!htbp]
\centering
\includegraphics[width=0.5\columnwidth]{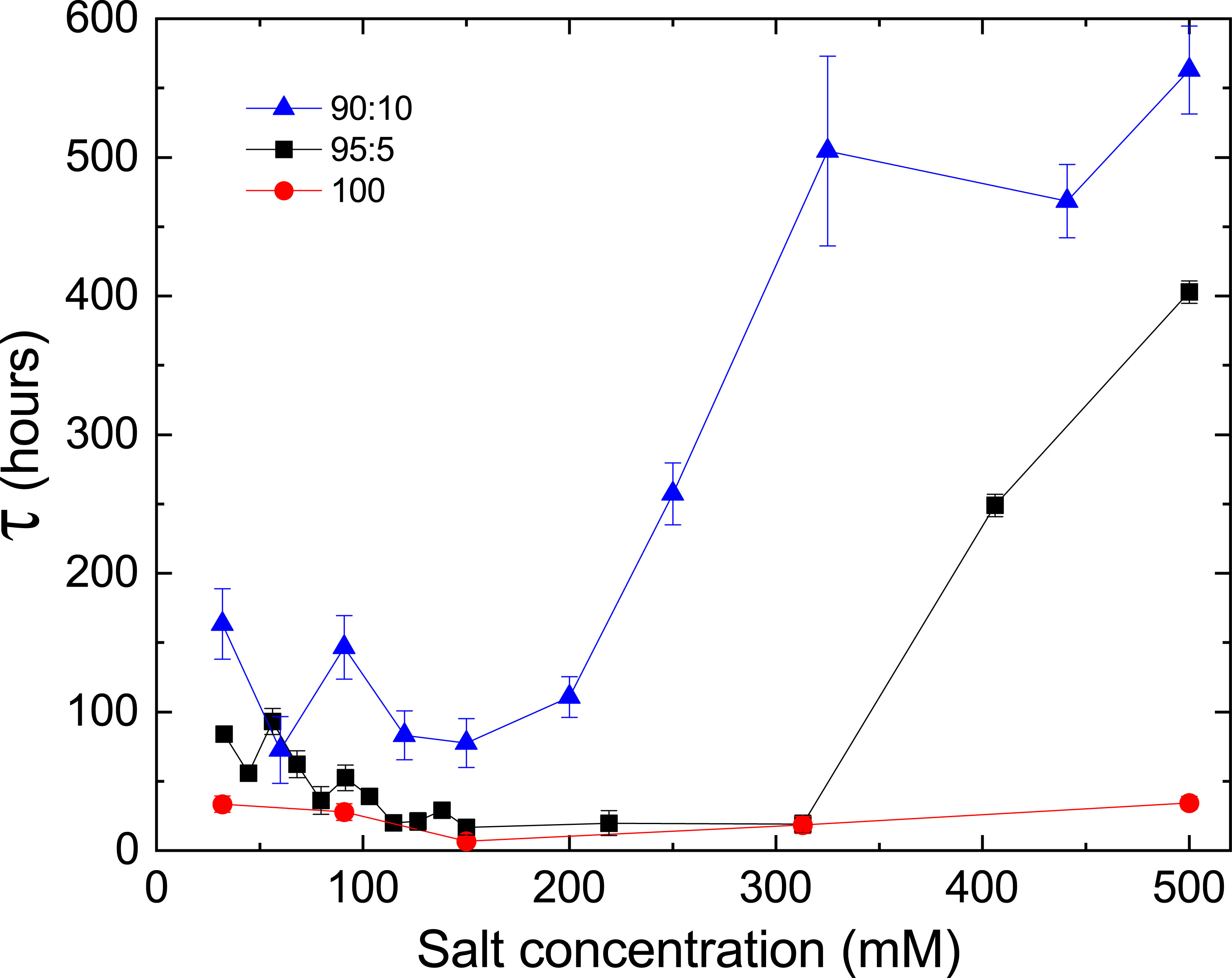}
\caption{Delay time as a function of monovalent salt concentration, at different DLPE:DLPG ratios (mole \%). At low salt concentrations (< 150 mM), samples that contain DLPG (black and blue) show an increase in the delay time as salt concentration is decreased. At high salt concentrations (>300 mM) the opposite trend occurs. However, when the sample contains pure DLPE, there is no significant effect of the salt concentration on the delay time.}
\label{fgr:salts}
\end{figure}

Since electrostatics are known to have a central role in stabilizing lipid lamellar systems, and the delay time is observed to increase with the fraction of charged PG lipids in our system, we tested the effect of the solution salt concentration on the delay time of samples with different DLPE:DLPG ratios. As shown in Fig. \ref{fgr:salts}, changing the average membrane charge density produces two features in the salt dependence: (a) there seems to be a minimum of the delay time at approximately 150 mM, splitting the dependency into two regimes, and (b) the delay time increases with the fraction of DLPG at a given salt concentration.

We propose that these two regimes originate from two different phenomena. At low salt concentrations (<150 mM), the decrease in delay time towards the minimum can be attributed to the decrease in the electrostatic screening length. At 150 mM the electrostatic screening length is comparable to the DLPG headgroup diameter ($\approx{} 8$ \AA{}) \cite{Pan2012}. Segregation of non-DLPE lipids, which is essential for recovering the homogeneous DLPE crystals, is facilitated by the screening of interacting charged PG headgroups. On the contrary, high salt concentrations (>300 mM) can lead to adsorption of ions on the charged membrane. This can lead to an increase in $\tau$, since ions must evacuate from between the lamellae, yet ion transport across membranes is unfavored. The adsorption of charges can result in an increase in the membrane's bending rigidity, which in turn can strengthen the metastability (see Sec. \ref{mechanism}).  The results show that the delay time of samples containing charged headgroups responded to changes in salt concentration, yet no significant dependence was observed in samples containing only DLPE. We would like to accentuate the extended lifetime of the metastable phase at the highest concentration measured (500 mM), which exceeded 500 hours in the case of 90:10 DLPE:DLPG (mole \%).

\begin{figure}[!htbp]
\centering
\includegraphics[width=0.5\columnwidth]{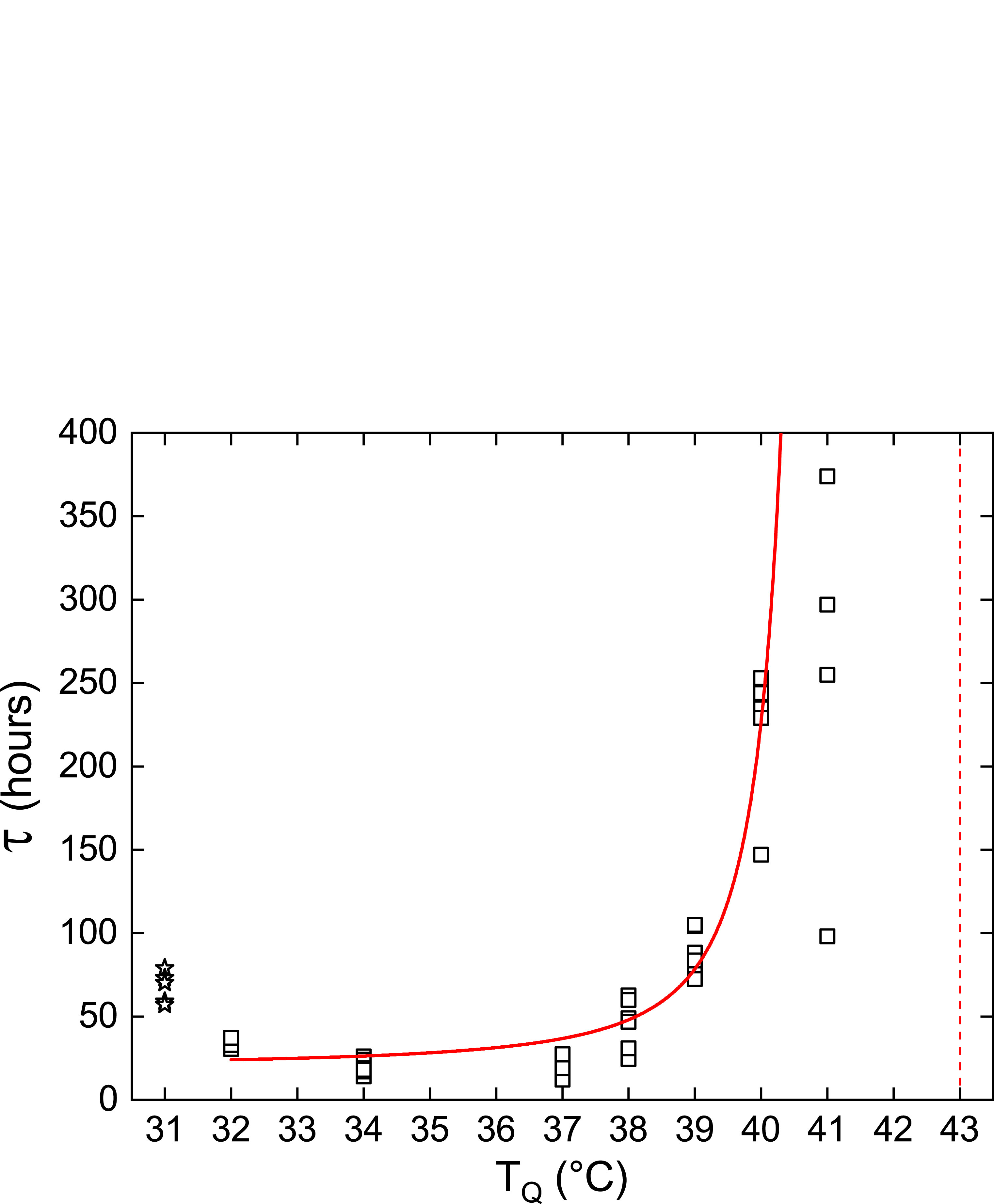}
\caption{Delay time as a function of quenching temperature. Empty squares represent measurements of individual capillaries at each quenching temperature. Solid red curve is the fit using Eq.~(\ref{tau}),  excluding the measurements at $T_{Q}=31$ \DC{} (empty stars represent the $L_\alpha \rightarrow L_\beta$ transition) and $T_{Q}=41$ \DC{} due to the high variance caused by fluctuations. Red vertical dashed-line indicates the estimated $T_{c\rightarrow\alpha}$ transition temperature.}
\label{fgr:TauTemp}
\end{figure}

The lifetime of the metastable phase depends on the strength of the thermodynamic force driving the transition. Close to the transition temperature the energy barrier that the system must overcome is high, which results in a small rate of nucleation. As the temperature is lowered, the barrier becomes smaller and the rate increases. This is demonstrated here by the increasing persistence of the metastable $L_{\alpha}$ phase, closer to $T_{c\rightarrow\alpha}$. Figure \ref{fgr:TauTemp} shows an exponential increase in the average delay time as a function of the quenching temperature, $T_{Q}$. Remarkably, for most quenching temperatures the spread of experimental results is very small, further supporting our claim that $\tau$ is an intrinsic property of the system's dynamics set by its macroscopic parameters. However, due to larger fluctuations closer to the critical temperature ($T_{c\rightarrow\alpha}$) we notice a large spread of $\tau$ at $T_{Q}=41$ \DC{}. There, two samples did not transition within the duration of the experiment (800 hours). In addition, at $T_{Q}=31$ \DC{} the $L_{\alpha}$ phase rapidly transitioned to $L_{\beta}$, which then became metastable. Evidently, the delay time for the $L_{\alpha}$ phase at $T_{Q}=32$ \DC{} is shorter than at $T_{Q}=31$ \DC{} for the $L_{\beta}$ gel phase, as the liquid phase is expected to be more labile \cite{Xu1988}.

The structural study of the delayed nucleation phenomenon, using time-resolved SXS, does not directly report on the thermodynamic processes. To address this, calorimetric measurements are commonly used to investigate the thermodynamics of lipid systems, mostly in the form of differential scanning calorimetry (DSC). However, since we are investigating a time-delayed transition at a fixed temperature, we employed DSC in a non-trivial quasi-isothermal manner. After samples were incubated at 60 \DC{} for 3 hours, the temperature was lowered to $T_{Q}=37$ \DC{}, and the samples were scanned back-and-forth between 36 and 37 \DC{} at a very slow rate (0.1 \DC{}/hour on heating, 0.43 \DC{}/hour on cooling).

\begin{figure}[!t]
\centering
\subfloat[]{\includegraphics[width=0.5\columnwidth]{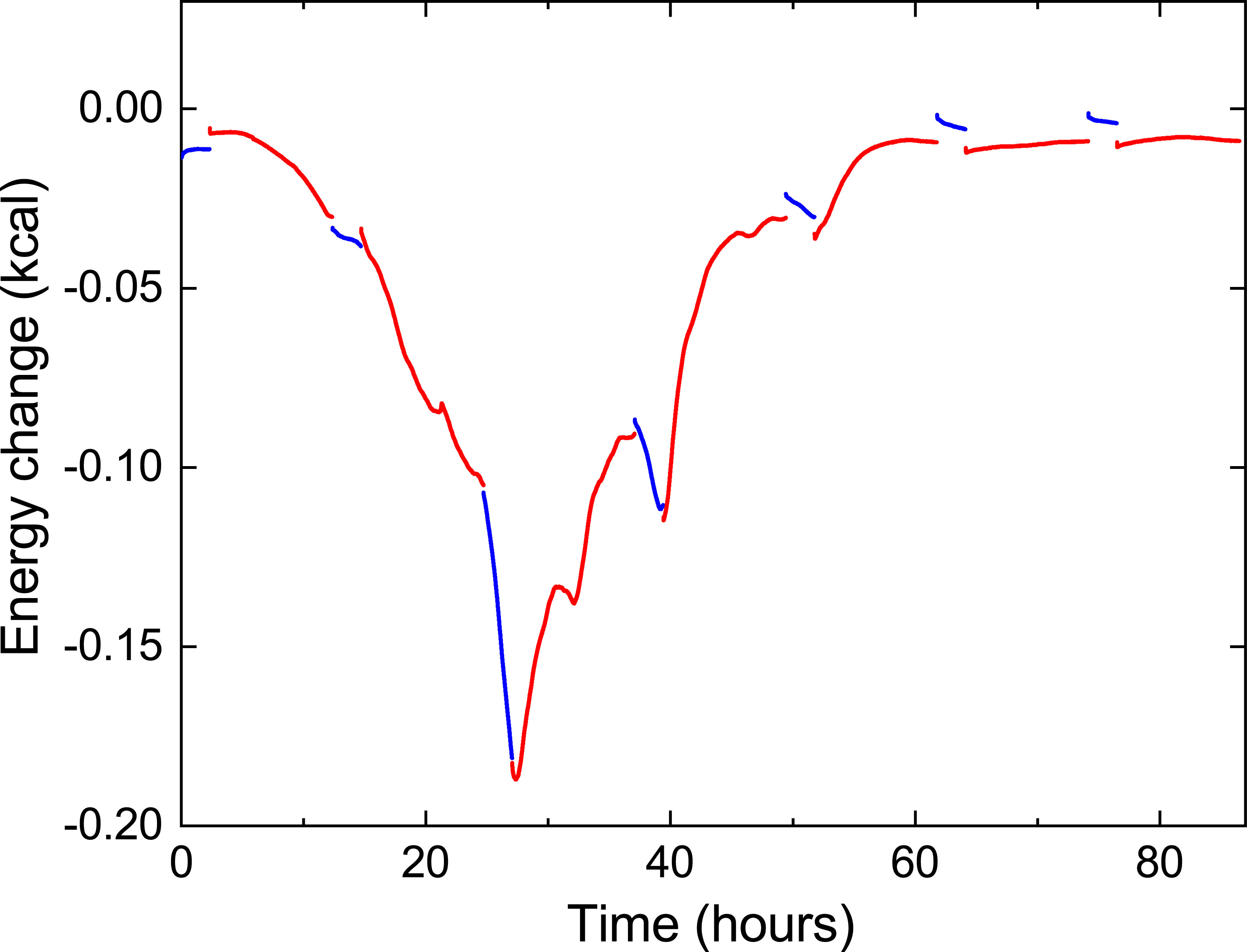}}
\subfloat[]{\includegraphics[width=0.5\columnwidth]{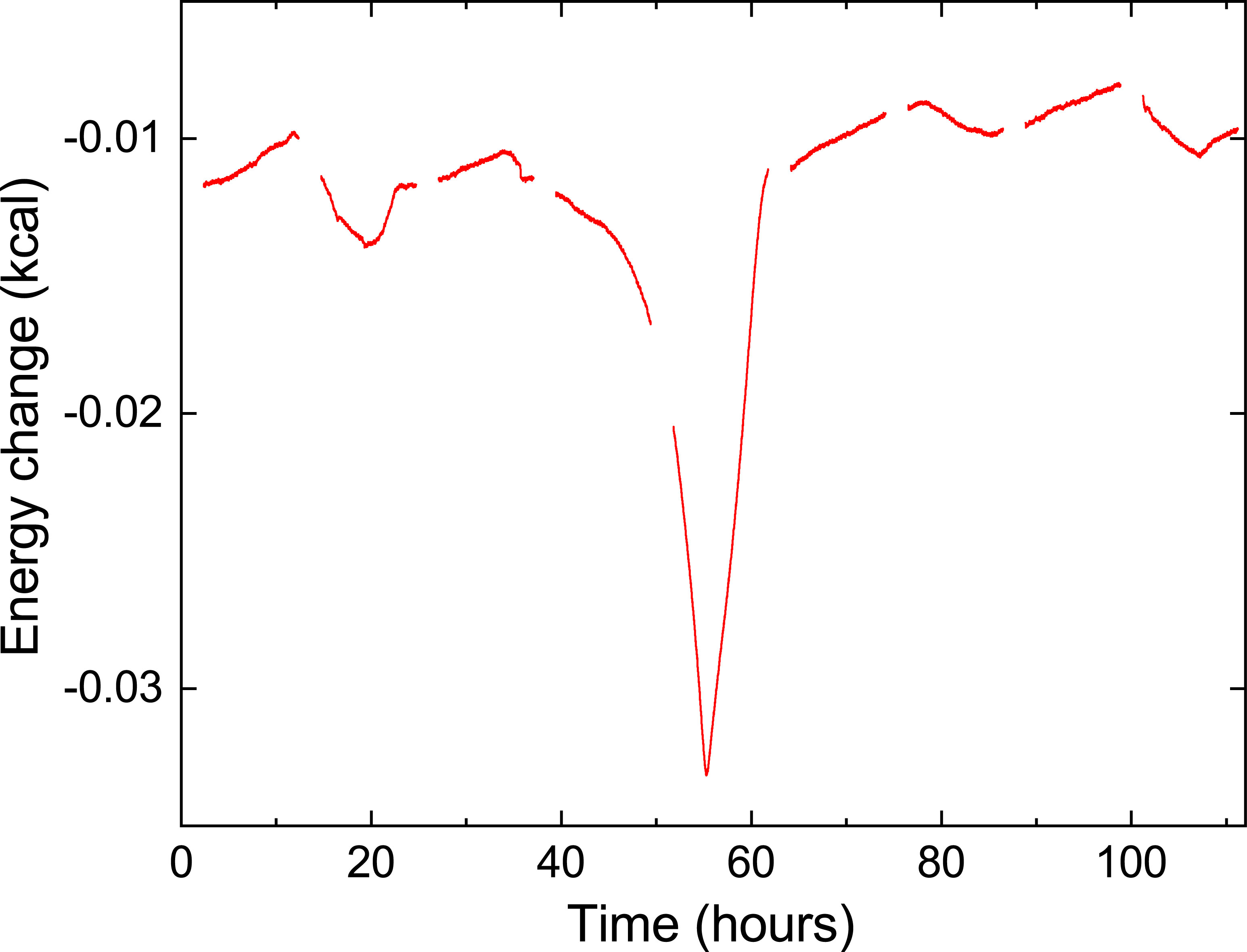}} 
\caption{Time-resolved quasi-isothermal DSC measurements of two samples: (a) Pure DLPE, (b) 90:10 DLPE:DLPG (mole \%). The exothermic peak seen in both occurs at similar times as in the scattering experiments. The red curves (shown in both) are heating scans and the blue curves shown only in (a) are cooling scans. There is a mismatch between cooling and heating scan data possibly due to the different scanning rates (see experimental section), thus the cooling scans were shifted in (a) and omitted from (b), for clarity. Raw data are shown in supplementary information.}
\label{fgr:DSC}
\end{figure}

In Fig. \ref{fgr:DSC} we compare the quasi-isothermal energy flux measurements performed on a sample of pure DLPE and a sample containing 90:10 DLPE:DLPG (mole \%). The results in both cases show an exothermic signature, as expected for the transition back to the low-energy crystalline phase, with similar timescales to those in our X-ray scattering measurements (Fig. \ref{fgr:PGPC}). The sample with pure DLPE shows a broad and slow change in the excess heat capacity, peaking at 28 hours, while the mixed sample shows a much narrower peak, centered at 55 hours. The corresponding average delay times in the X-ray scattering experiments are 20 and 45 hours, respectively. In addition, a DSC measurement was performed to determine the enthalpy of transition from the $L_c$ to the $L_\alpha$ phase for a pure sample of DLPE, which yielded $h_{\rm c}=11.1$~kcal/mole.

\begin{figure}[!htbp]
\centering
\subfloat[]{\includegraphics[width=0.5\columnwidth]{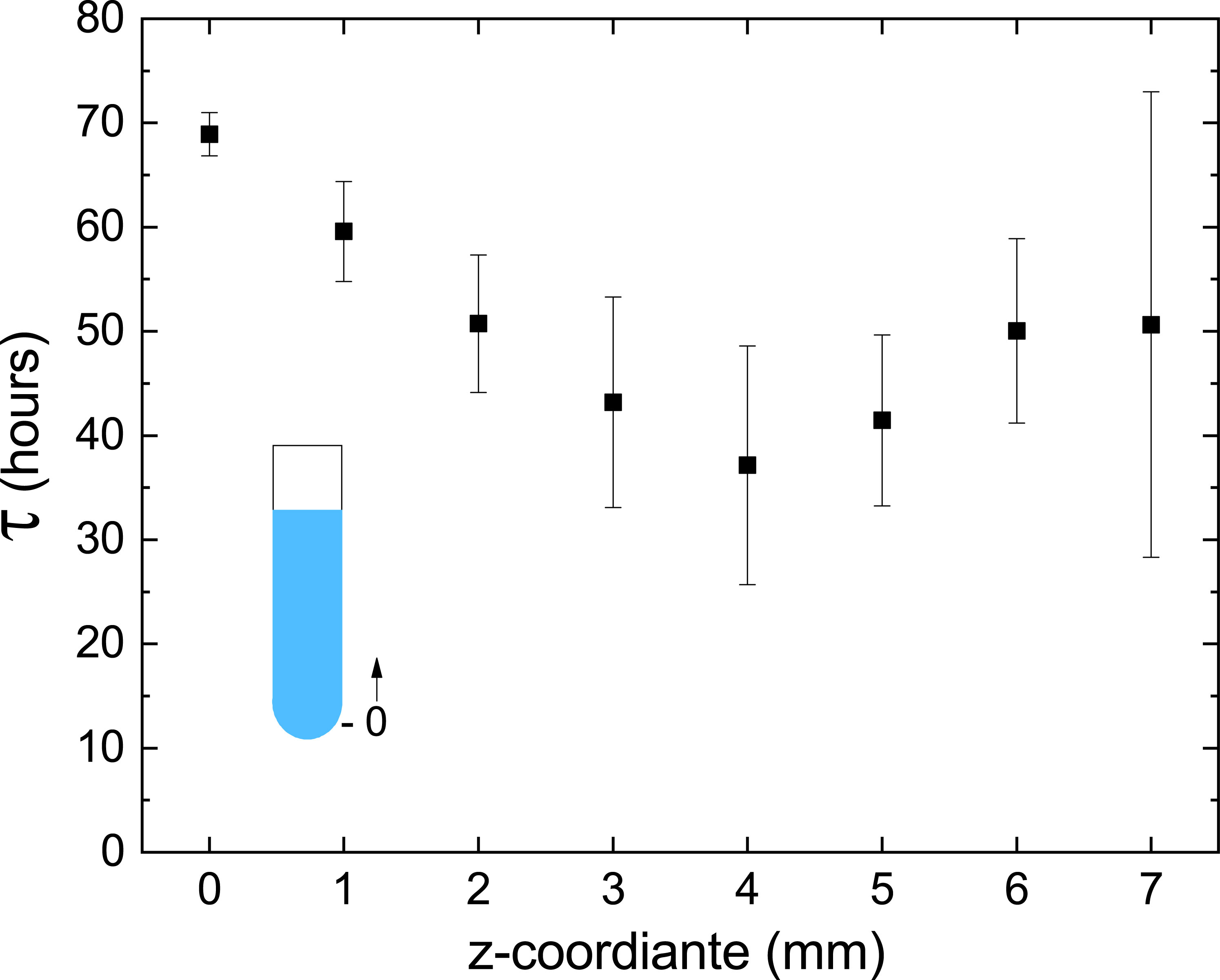}}
\subfloat[]{\includegraphics[width=0.5\columnwidth]{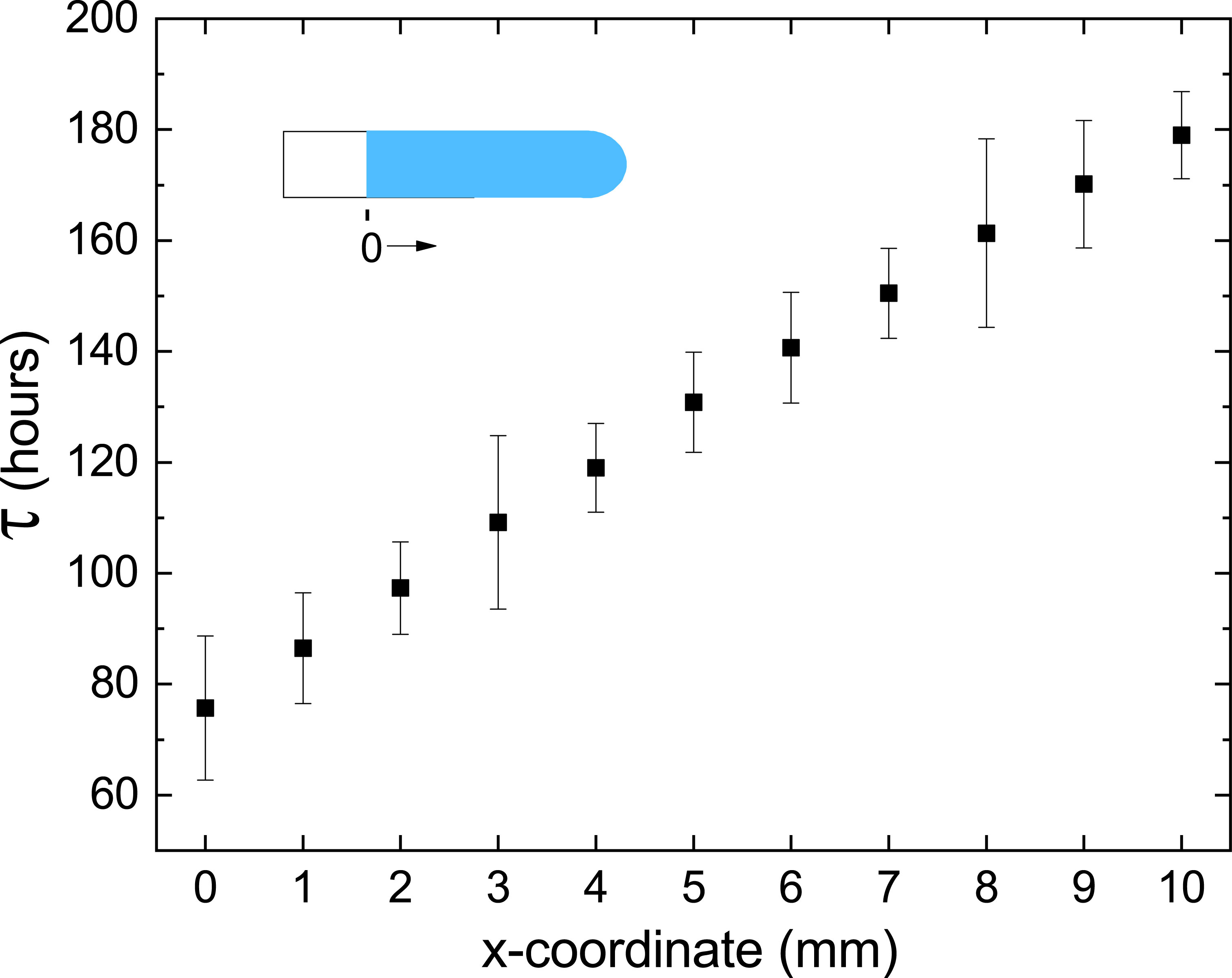}}\\
\subfloat[]{\includegraphics[width=0.5\columnwidth]{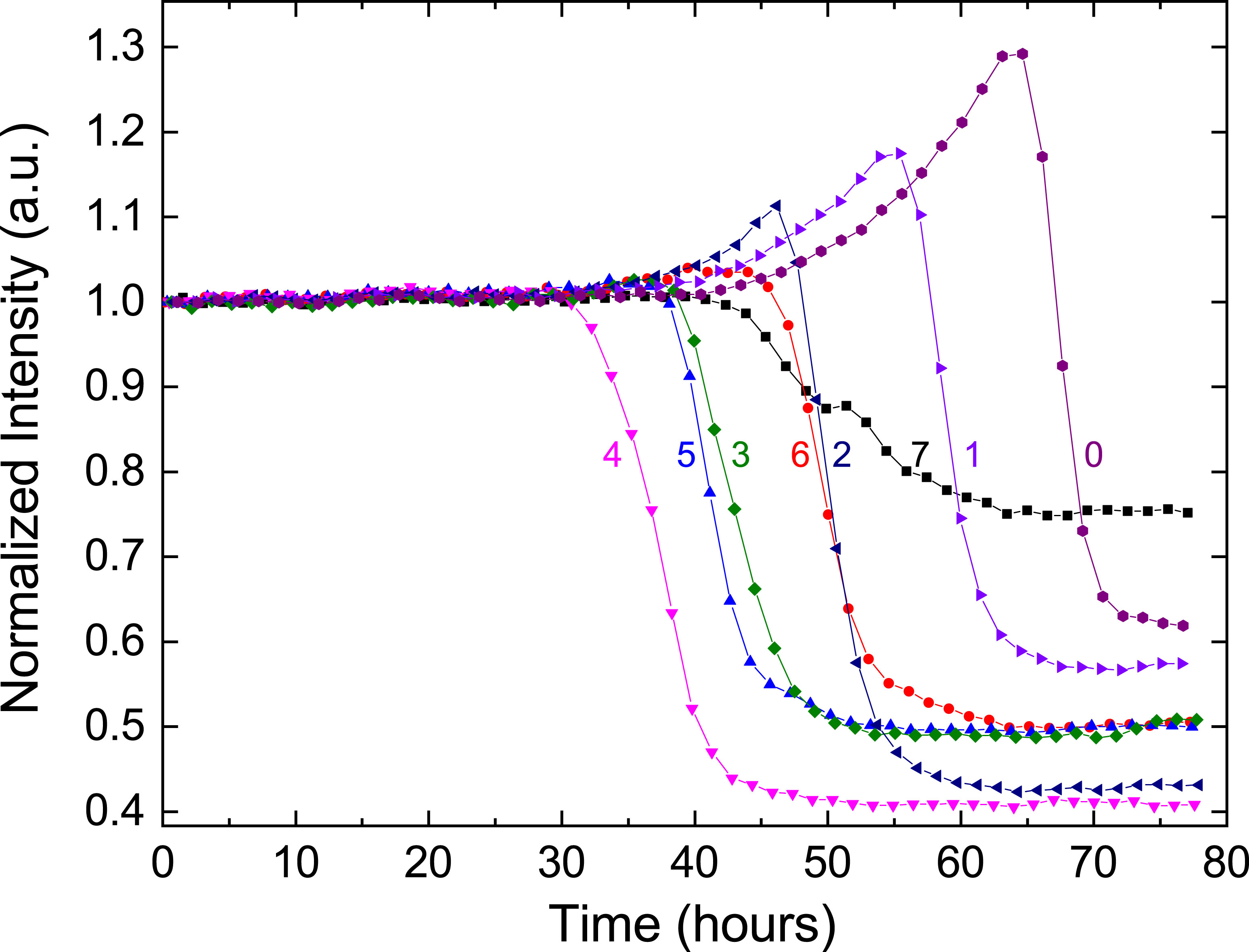}}
\subfloat[]{\includegraphics[width=0.5\columnwidth]{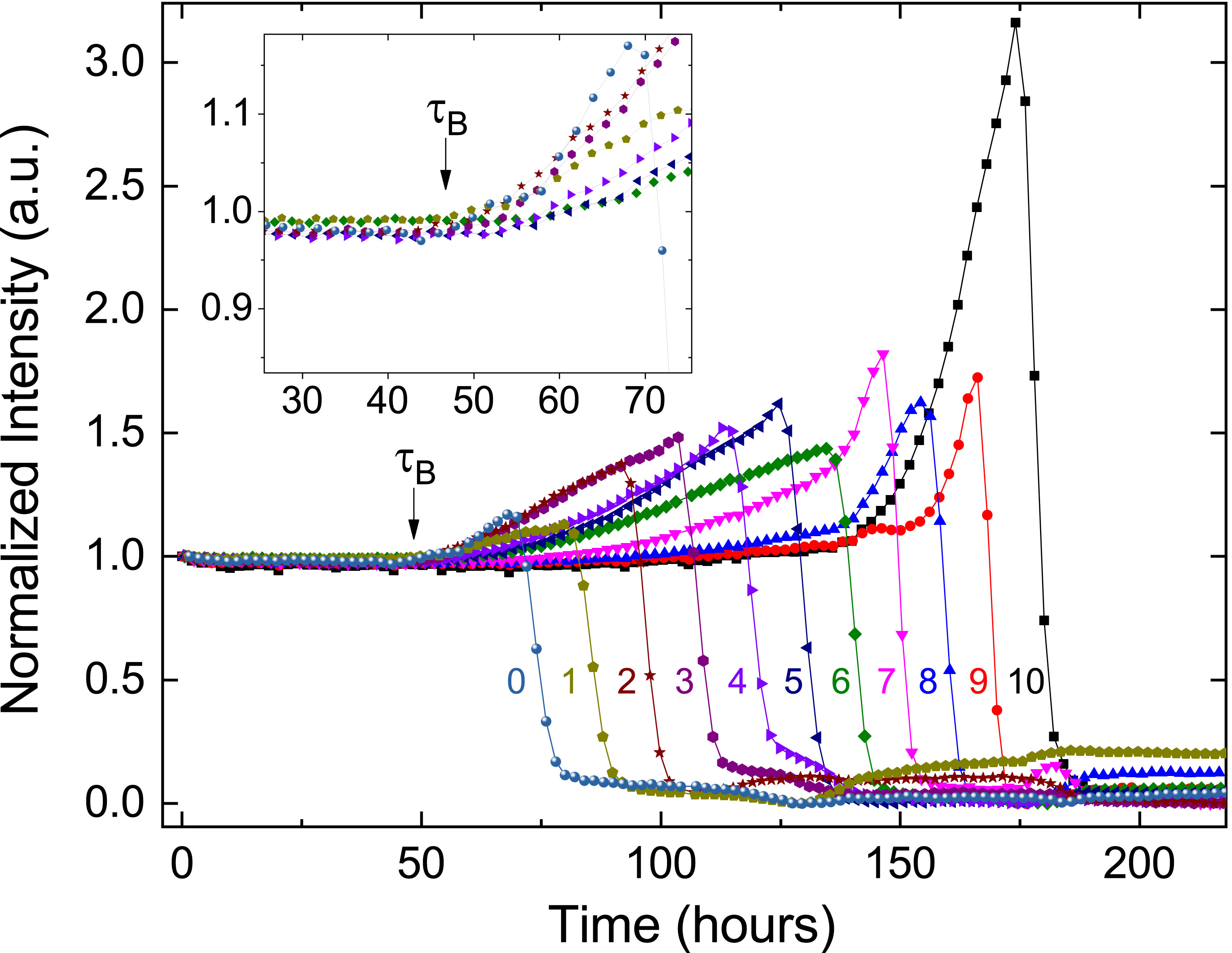}}
\caption{The delay time occurs at different times for different locations, but structural changes to the lipid particles are coordinated over millimeters. (a) Delay time vs. spatial coordinate in vertically held capillary. (b) $\tau$ vs. spatial coordinate in horizontally held capillary. (c) Time-resolved scattering intensity of (001) as a function of time. Error bars in (a) and (b) represent $\tau^{*}$, the duration of the transition. Numeric tags in (c) and (d) correspond to spatial coordinate in (a) and (b), respectively. Inset in (d) shows $\tau_{B}$, the beginning of the build up period.}
\label{fgr:TauCoordinates}
\end{figure}

In the process of sample measurement and analysis of the phase-transition dynamics, the extracted temporal parameters represent the transformation occurring in the illuminated volume. However, this volume includes only a portion of the lipid pellet at the bottom of the capillary. If one follows a nucleation and growth framework, it is important to assess whether the transformation initiates concurrently throughout the sample or propagates successively from a starting point. To test this, we prepared a sample with a large pellet at the bottom of the capillary ($\approx 10~mm$), and measured the delay time at different locations along the vertically held capillary. The capillary size and experimental protocol (SXS) remained as previously described. Figure \ref{fgr:TauCoordinates}a shows the delay time as a function of the spatial coordinate along the capillary. The transition seems to propagate outward from a certain location, with neighboring locations transitioning at later times.

A similar experiment was performed on a horizontally held capillary. There, the transition began at the water-pellet interface and propagated at a steady velocity of approximately 100 $\mu$m/hour towards the end of the pellet (Fig. \ref{fgr:TauCoordinates}b). In both experiments a new feature in the scattering spectra could be observed: a coordinated increase in lamellar scattering over a period of time prior to the transition (Fig. \ref{fgr:TauCoordinates}c, d). We denote the beginning of this period by $\tau_{B}$, the point in time from which a slow increase in scattering culminates in a sharp drop of intensity. Only after the drop in the lamellar scattering intensity is there a detectable change in wide-angle scattering. Therefore, the metastable state remains during the build-up period. Surprisingly, this structural reorganization is coordinated over several millimeters in the sample (Fig. \ref{fgr:TauCoordinates}c, d).
 
\begin{figure}[!t]
\centering
\includegraphics[width=0.5\columnwidth]{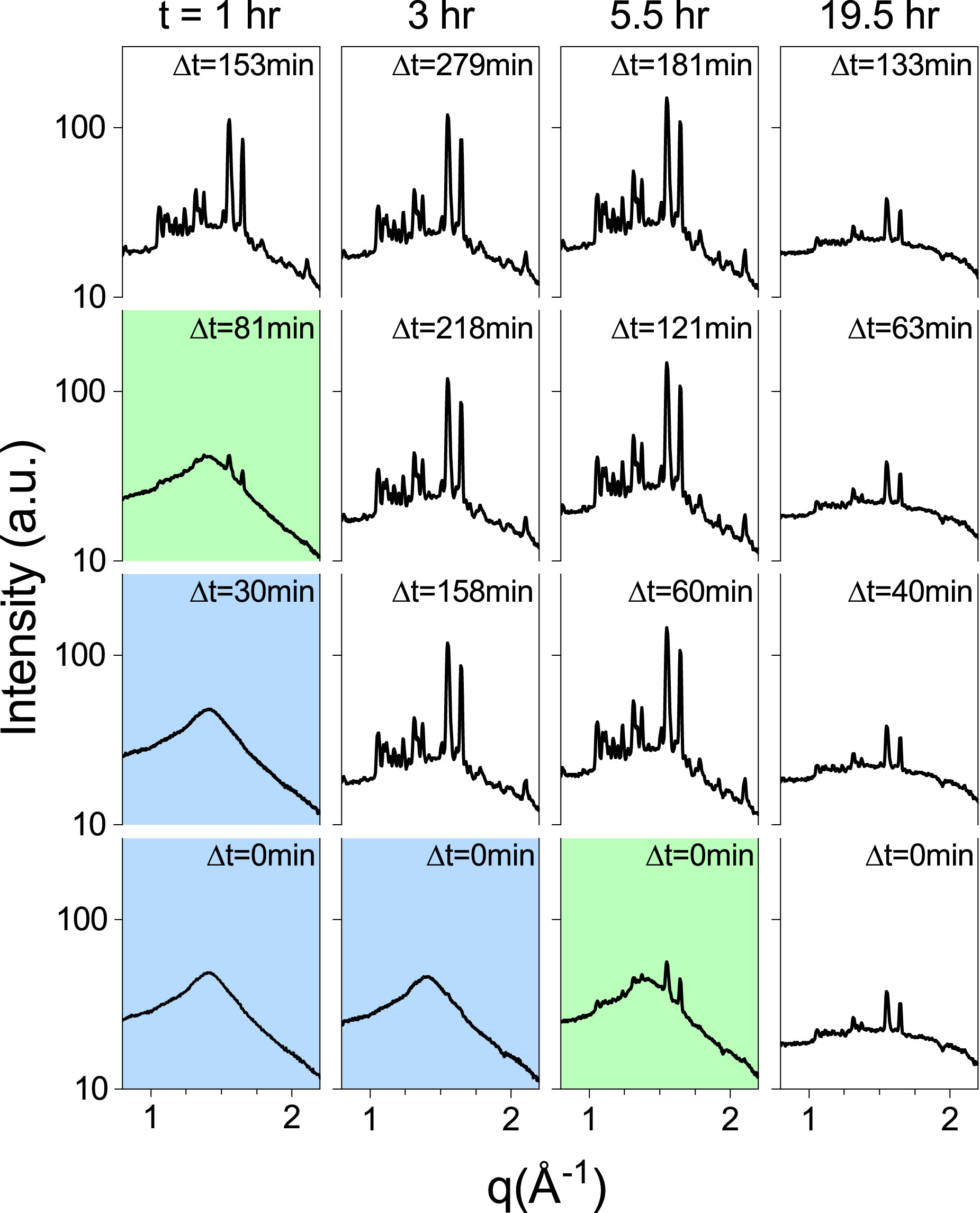}
\caption{The metastable state's lifetime is significantly shortened by applying mechanical agitation in the form of rigorous pipetting. From left column to right, samples that were taken from a bulk reservoir $t$ = 1, 3, 5.5 and 19.5 hours after temperature quenching from $T=60$ \DC{} to $T_Q=37$ \DC{}. Each sample is subjected to pipetting, and measured immediately after extraction and $\Delta t$ minutes afterwards. Blue background indicates the sample is still in the metastable $L_\alpha$ phase, and green that the sample is mid-transition.}
\label{fgr:Eppendorf}
\end{figure}

Metastable phases are often very sensitive to energy fluctuations, as even minute inputs of energy can result in a transition to the stable phase. Since the lipid metastable phase is stable against various changes in system parameters, we tested its stability against external inputs of energy by subjecting lipid dispersions to mechanical agitation in the form of rigorous pipetting. A lipid dispersion of approx. 1.5 ml, at 30 mg/ml, was prepared as a bulk dispersion from which samples would be pipetted out, and measured intermittently. It was incubated at 37 \DC{} for one hour, followed by 3 hours at 60 \DC{}, as performed regularly with the SXS samples. The incubator was then set to $T_{Q}=37$ \DC{} and a sample was drawn from the bulk dispersion after t = 1, 2, 3, 5.5 and 19.5 hours by pipetting out approximately 100 $\mu l$ and placing into a capillary. The capillary was then placed in the SXS temperature chamber, pre-heated to 37 \DC{}, and measured after $\Delta t$ minutes (Fig. \ref{fgr:Eppendorf}). The control sample, taken from the bulk dispersion before it was placed in the incubator, underwent the regular temperature procedure in the SXS temperature chamber, and transitioned after $\tau=19.5$ hours. The samples from the bulk dispersion, taken during first few hours, transitioned approximately an hour after being pipetted out. The sample taken after 5.5 hours was in the middle of transitioning when measured initially ($\Delta t=0~min$), and the 19.5 hour sample had already transitioned (Fig. \ref{fgr:Eppendorf}). This experiment demonstrates that the lifetime of the metastable phase is significantly shortened by mechanical agitation applied after thermal incubation.

\begin{figure}[!htbp]
\centering
\includegraphics[width=0.5\columnwidth]{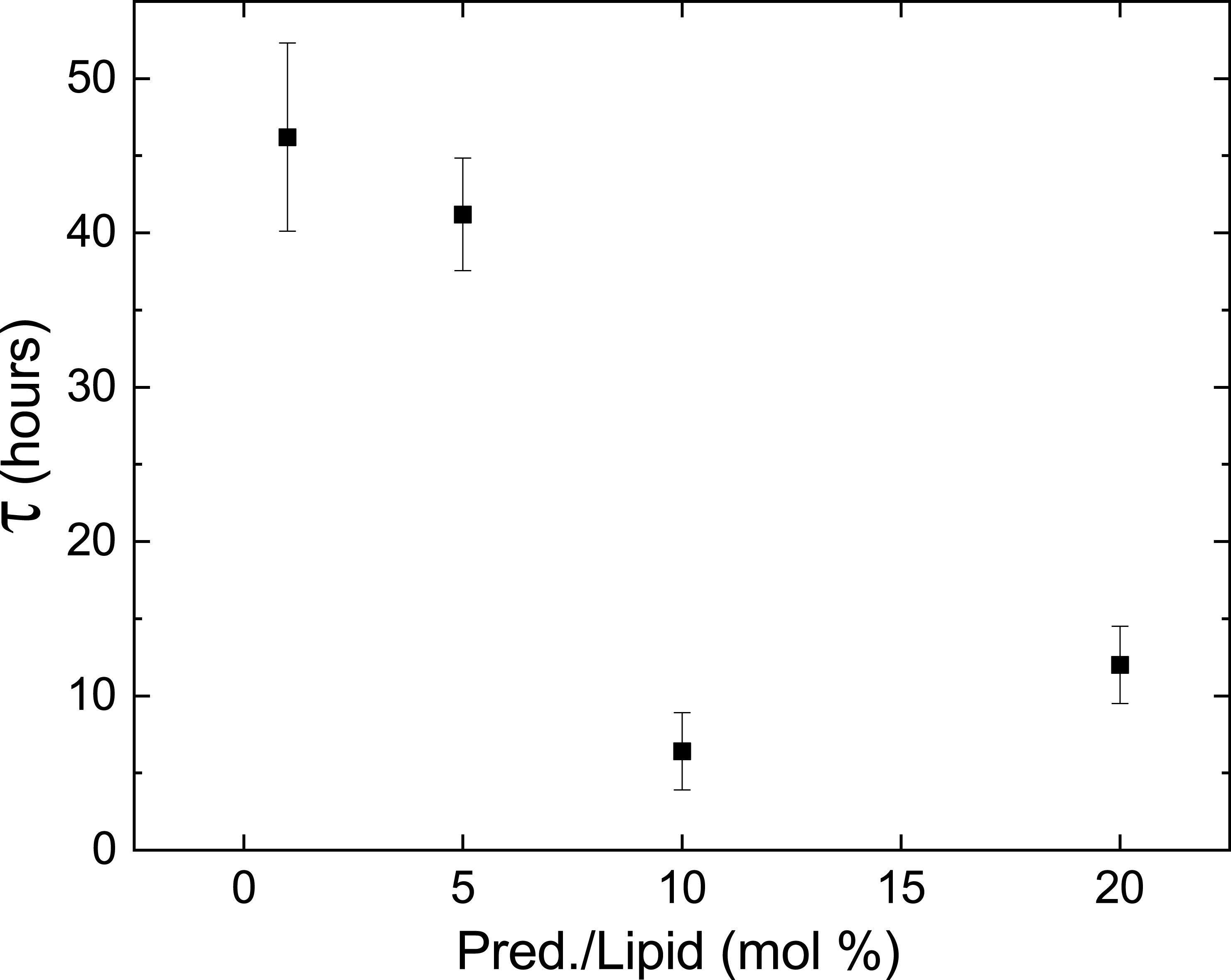}
\caption{The delay time decreases in the presence of prednisolone, a commercial hydrophobic drug, but the delay in transition still persists.}
\label{fgr:prednisolone}
\end{figure}

Lastly, phospholipids are utilized in bio-medicine as building blocks of vesicles designed for specific targeting and controlled release. When designing such drug-delivery systems, it is crucial to assert the stability of the carrier with its cargo. Since DLPE and DLPG have been used as the lipid components of such systems \cite{Cohen2014, Rivkin2010, Bachar2011}, we tested the stability of the metastable phase in the presence of cargo. Figure \ref{fgr:prednisolone} shows the delay time of samples containing 90:10 DLPE:DLPG (mole \%) and the hydrophobic drug Prednisolone; an established and commercially available steroid used to treat a wide range of conditions and illnesses. The results show that the addition of the drug had a large impact on the delay time, shortening it by almost an order-of-magnitude. Since lipid systems continue to serve as appealing ingredients in drug delivery systems, controlled delayed nucleation may serve as a novel designing factor to deposit cargo in a predetermined timing. Nonetheless, the effect of cargo on transition dynamics should not be overlooked when designing such systems.

\section{Discussion}
\subsection{Experimental results}
\begin{figure}[!htbp]
\centering
\includegraphics[width=0.5\columnwidth]{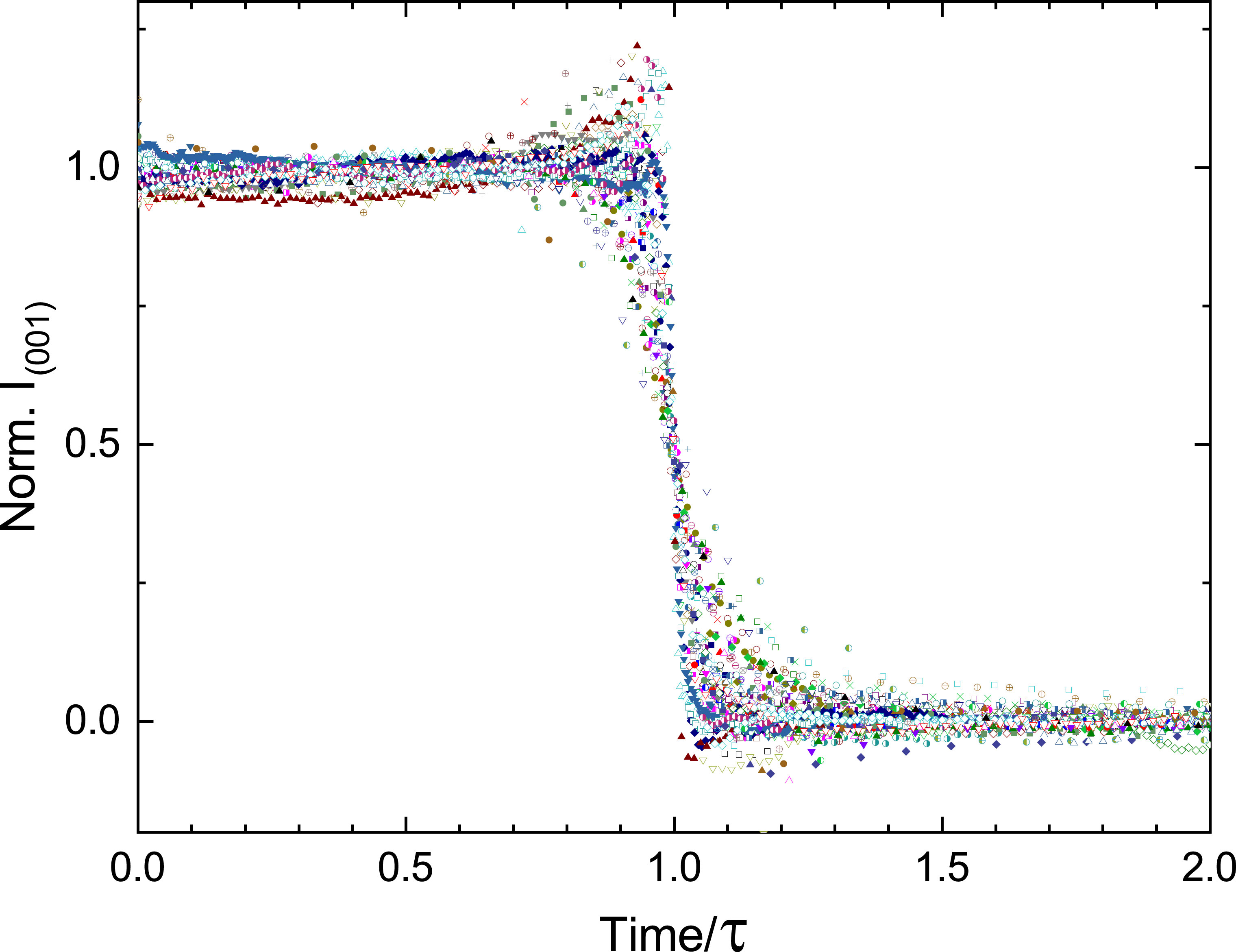}
\caption{The lamellar correlation peak intensity over time can be re-scaled by $\tau$ to highlight the cooperative nature of the transition, regardless of the conditions changed in the experiments. The residual variation in the time-dependent scattering curves is due to $\tau^*$. The data set presented here consists of 85 different experiments.}
\label{fgr:collapse}
\end{figure}

The findings presented in this work are in clear contrast to those expected from a system described by CNT. Instead of a single stochastic process, which would produce a single timescale for the transition, we present multiple experimental evidence of coordinated delayed nucleation and multiple timescales ($\tau$, $\tau^*$ and $\tau_{B}$), representing the complexity of the dynamics. These timescales are orders of magnitude larger than the typical microscopic timescales associated with lipid systems. To emphasize the separation of timescales and highlight the collective behavior of the transition, we re-scale our entire data set of lamellar-scattering peak intensities by the delay time $\tau$ (Fig. \ref{fgr:collapse}). The time-dependent intensities collapse onto roughly the same sigmoidal shape, with slight variations in width representing the variations in $\tau^*$.

We present additional findings that we would like to discuss in the context of deviations from CNT. Firstly, it is important to mention that at no point in the preparation of the samples was there any effort to homogenise the particles' sizes. And yet, despite this heterogeneity, the delay time was shown to be reproducible with a peak in the probability at a non-zero value \cite{Jacoby2015}. Secondly, our results shown in Fig. \ref{fgr:TauCoordinates} demonstrate that the transition occurs at different times at different locations, yet the onset of the structural change in samples with large pellets, marked by $\tau_B$, is macroscopically coordinated over millimeters. This length-scale is orders of magnitude larger than any microscopic length-scale associated with lipid self-assembly. Lastly, in heterogeneous nucleation impurities are considered preferential nucleation sites due to a lower surface energy penalty compared to the homogeneous case. In our system, the inclusion of a secondary lipid only served to hinder crystallization.

During the incubation time at the high temperature ($\geq 60$ \DC{}) water molecules and ions enter in between the membranes \cite{Jacoby2015}. Concurrently, secondary lipid molecules enter the liquid membrane and disrupt its homogeneity. Upon cooling, the water molecules and ions must evacuate, and lipid segregation must occur to re-form the network of connections as in the initial homogeneous crystal. Here, we demonstrated that the persistence of the metastable phase is sensitive to the properties of the secondary lipid. Not only does the metastability with its features persist, but the chemical structure of the lipid has a large impact on the change of the delay time (Fig. \ref{fgr:PGPC}).

A recent study experimentally showed long-range interlayer alignment of phase-separated intralayer domains, across hundreds of lamellae in multi-component supported lipid membranes \cite{Tayebi2012}. A follow-up study proposed a theoretical explanation to the interlayer correlation between phase-separated domains \cite{Hoshino2015}. Using a model of stacked 2D Ising spins to represent the stacked lipid membranes, they showed that the system forms a continuous columnar structure in equilibrium, for any finite interaction across adjacent layers. Such an interlayer interaction should be a key component in cooperative nucleation in MLVs, for which a mechanism is proposed below.

\subsection{Metastability mechanism} \label{mechanism}
We would like now to examine a possible mechanism for the exceptional metastability of the $L_\alpha$ phase and the strong cooperativity of the transition. 

Because of the rigidity of the crystalline phase, the formation of a crystalline domain in a membrane flattens that region, which affects adjacent membranes and thus locally deforms the MLV. The free-energy penalty per unit area due to the deformation is proportional to the effective surface tension $\gamma=\sqrt{BK}$ of the MLV, arising from its compression ($B$) and bending ($K$) moduli \cite{deGennes1993}. This penalty makes the free-energy of the crystalline phase effectively higher, shifting the transition from $T^{(0)}_{\rm c}$ for an isolated membrane to a lower temperature $T_{\rm c}$ for a membrane in a curved MLV. The change in the free-energy per unit area between the two phases is $\Delta g(T_{\rm c}) \simeq 2\gamma/3$. Using the relation $\Delta g \simeq (h_{\rm c}/a) (T^{(0)}_{\rm c}-T)/T^{(0)}_{\rm c}$, the temperature shift can be estimated from the measured enthalpy of transition ($h_{\rm c}=11.1$~kcal/mole), the area per lipid ($a \simeq 0.5$~nm$^2$), and assuming $\gamma$ larger than $0.1$ mN/m. We get a decrease in the transition temperature, proportional to $\gamma$, of more than 10 degrees.

Thus, under conditions where an isolated membrane would crystallize, a single membrane in the MLV would not. On the other hand, if all the membranes in the MLV were to crystallize, the total free-energy would inevitably decrease. Hence, the MLV must ultimately crystallize, but it can do so only through a multi-membrane cooperative process. This cooperativity is essential for departing from a single Poisson process, typical to CNT.

The compression and/or bending moduli of the $L_\alpha$ phase, and therefore also $\gamma$, increase with increasing lipid chain length and increasing membrane charge. Such modifications, by increasing $\gamma$, should deepen the metastability. This is consistent with the experimental observations reported here.

Let us consider the collective metastability described above in slightly more detail. Since the deformation of the membrane stack is localized \cite{deGennes1993}, its free-energy penalty is intensive in the number of membranes, whereas the free-energy gain due to crystallization is extensive. Hence, there is a critical number of membranes, $n_{\rm c} \sim \gamma/(\Delta g)$, beyond which the suppression of the transition is overcome. The multi-membrane critical nucleus has the size $R_{\rm c} \sim \lambda n_{\rm c}$, where $\lambda$ is the MLV's periodicity. It corresponds to a multi-membrane nucleation barrier, $F_{\rm c} \sim \sigma R_{\rm c}^2 \sim \sigma\gamma^2\lambda^2/(\Delta g)^2$, where $\sigma$ is an effective surface tension; a combination of $\gamma$ and the line tension of intra-membrane crystalline domains.

To sum up this analysis, we obtain the multi-membrane nucleation barrier as
\begin{equation}
  F_{\rm c} \sim \frac{\bar\sigma^3 \lambda^2 a^2}{h_{\rm c}^2} \left( \frac{T_{\rm c}}{T_{\rm c}-T}
  \right)^2,
\label{Fc}
\end{equation}
where $\bar\sigma=(\sigma\gamma^2)^{1/3}$ is an effective surface tension. This implies a delay time of the form
\begin{equation}
  \tau = \tau_0 e^{F_{\rm c}/(k_{\rm B}T)} = \tau_0 e^{b/(T_{\rm c}-T)^2},
\label{tau}
\end{equation}
where the coefficient $b$ is extracted from Eq.~(\ref{Fc}), and $\tau_0$ will depend on faster effects not addressed here. Using the experimental values of $\lambda$, $a$, $h_{\rm c}$, and $T_{\rm c}$, and taking $\bar\sigma\sim 1$~mN/m, we get $b$ of order $10$ K$^2$. This value is sensitive to the value of $\bar\sigma$ and should be regarded just as a consistency check.

Figure \ref{fgr:TauTemp} shows the fit of the function in Eq.~(\ref{tau}) to the data, where $T_{\rm c}$ was fixed to 43 \DC{}. The fitting parameter $b$ yields a value of $\approx 22$ K$^2$, consistent with the qualitative estimate shown above. Changing $T_{\rm c}$ to be 42 or 44 \DC{} changes $b$ to be approx. 8 or 44, respectively. A detailed theory along the lines presented in this section will be presented elsewhere.

\section{Conclusions}
Our results herein show a deterministic and controllable behavior of the lipid metastable phase. Our experimental results demonstrate the sensitivity of the delay time to a variety of system parameters. Furthermore, the long timescales and robustness against changes allow one to access the complex dynamics, which could otherwise be a difficult obstacle to overcome. Finally, the proposed mechanism is consistent with the experimental results and might account for the cooperative nature of the transition. A better understanding of the mechanism for delayed nucleation can benefit both the fundamental physics of nucleation processes and applications harnessing it for releasing cargo in a timely manner.

\bibliographystyle{unsrt}

\end{document}